\documentclass[aps,prl,epsfig,floats,twocolumn,amssymb,amsmath,showpacs,footinbib]{revtex4-1}

\usepackage{color}

\usepackage{graphicx}

\begin{document}

\title{Reply to Comment on ``Length Scale Dependence of DNA Mechanical Properties''}

\author{Agnes Noy}
\affiliation{Rudolf Peierls Centre for Theoretical Physics, University of Oxford, 1 Keble Road, Oxford, OX1 3NP, UK}
\author{Ramin Golestanian}
\affiliation{Rudolf Peierls Centre for Theoretical Physics, University of Oxford, 1 Keble Road, Oxford, OX1 3NP, UK}

\date{\today}


\maketitle

\begin{figure}[b]
\includegraphics[width=0.90 \columnwidth]{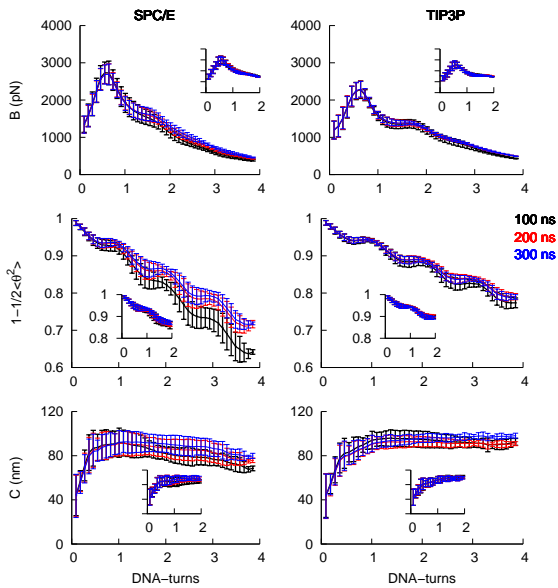}
\caption{Length scale dependence of stretch modulus B (Top), directional decay (Middle) and Torsional persistence length C (bottom) evaluated from different trajectories lengths using the 56mer (inset 36mer) fragment}\label{figure1}
\end{figure}

In a Comment \cite{mazur-comment} on our recent study of the length scale dependence
of DNA elasticity \cite{noyPRL12}, Mazur brings up a number of technical points about
the analysis of the MD trajectories used in our simulations, broadly categorized
as concerns about equilibration and interpretation of the observed results.
In Ref. \cite{noyPRL12}, we compare the results we obtain for the segment with 4 DNA
turns to that with 2 DNA turns, and interpret their agreement as a sign of equilibration.
The shorter segment (2 DNA turns), which we simulated and analyzed for 130 ns, safely
falls within the time scale of equilibration, and so does most of the analysis for
the longer segment (4 turns), with the possible exception of the bending mode which
might be marginal. We have been careful to avoid drawing conclusions only based on
the information at the longest length scale. For instance, twist or contour-length
elastic profiles from both fragments present the plateau at lengths shorter than
a DNA turn, and bend modulations are uniformly distributed at all studied lengths.
The only long-scale effect reported in our paper was the end-stretching mode,
which is already present for the shorter segment, and does not directly involve
bending. The choice of water model should not affect the overall equilibrium
properties \cite{noy09}, but will affect relaxation times and could also incur quantitative
differences \cite{note}.

We have performed new simulations and produced 300 ns trajectories using both water
models, to further probe the issue of equilibration (see Fig. \ref{figure1}).
Figure \ref{figure1} shows that the 36mer fragment is clearly equilibrated, and
that the two water models lead to the same qualitative features, although with
a quantitative difference in the rate of bending decay (i.e. value of persistence length).
The 56mer fragment with the TIP3P model is equilibrated, and so is the twist and
stretch response of the SPC/E case, while the bending shows small differences
between 100 ns, 200 ns, and 300 ns trajectories \cite{note2}.

When we start from an atomistic resolution, the definition
of the DNA axis on the bp-level can be somewhat arbitrary.
However, with the exception of rise, the influence of either
the choice of the reference frame or the mathematical algorithm
is shown to be negligible \cite{frame}. Similar bending periodicity to what
we report has been observed using FREEHELIX \cite{dickerson97},
CURVES \cite{mazurJPCB09} and Monte Carlo coarse-grained simulations
over long molecules of DNA \cite{Olson88}. Moreover, Dickerson and
coworkers described the same helical periodicity analyzing entirely
experimental structures \cite{dickerson97}. This effect is due entirely
to the static structure of DNA, namely its spontaneous curvature.
Because DNA is a twisted polymer, this periodicity emerges when
a regular DNA with a systematic positive roll is built (blue line of
Fig. 3c \cite{noyPRL12}). Due to this residual average static (helical)
structure, orientation correlation function will have a periodic shape
even at zero temperature and without any fluctuations. If one artificially 
uses an axis that averages over this static structure, one can eliminate 
the periodicity. What we point out about definitions of persistence length 
from the decay of this function is aimed at alerting to the fact that using 
a local slope, as has been used in interpreting some experimental data, 
might give erroneous results due to this residual structure.

For the reasons given above, we do not believe the Comment
by Mazur provides any evidence against our work.


\begin{thebibliography}{99}
\bibitem{note}
The SPC/E water model gives the best bulk water dynamics and structure
compared with experiments, including self-diffusion coefficients and radial distribution
functions (RDF) between O-O, H-O and H-H water atoms. On the other hand, TIP3P
is less structured lacking the second peak of the O-O RDF \cite{01nilsson}.
However, the free energies of solvation obtained with TIP3P give a closer
match to experiments, whereas those from SPC/E are too low \cite{05pande}.
\end{thebibliography}
\end{document}